\documentclass{article}
\usepackage{styles/spconf}

\usepackage{algorithm}
\usepackage[noend]{algorithmic}

\usepackage{cite}
\usepackage{amsmath,amssymb,amsfonts}
\usepackage{graphicx}
\usepackage{textcomp}
\usepackage[svgnames,x11names]{xcolor}

\usepackage{lipsum}

\usepackage{array}
\usepackage{mathtools}
\usepackage{tcolorbox}
\usepackage{color}
\usepackage{theorem}
\usepackage{amssymb}
\usepackage{cite,hyperref}
\usepackage{cases}
\usepackage{url}
\usepackage{enumitem}
\usepackage{multirow}
\usepackage{hhline}

\input{styles/mysymbol.sty}

\usepackage{tikz}
\usetikzlibrary{shapes,arrows}
\usepackage{moresize}
\usepackage{pgfplots}
\usepackage{pgfplotstable}
\usepackage{wrapfig}
\usepgfplotslibrary{groupplots}
\usepgfplotslibrary{fillbetween}
\pgfplotsset{compat=1.17}
\pgfplotstableset{col sep=comma}

\newtcolorbox{myblockt}[1]{colback=urblue!5!white,
	colframe=urblue,fonttitle=\bfseries,
	title=#1}
\newtcolorbox{myblock}{colback=urblue!5!white,
	colframe=urblue,fonttitle=\bfseries}

\pgfplotsset{compat=1.17}
\pgfplotstableset{col sep=comma}
\tikzset{every mark/.append style={scale=1.5, solid}, font=\footnotesize}
\pgfplotsset{
    width=1.05\textwidth,
    legend style={
        font=\scriptsize ,  
        inner xsep=1pt,
        inner ysep=1pt,
        nodes={inner sep=1pt}},
    legend cell align=left,
	every axis/.append style={line width=0.5pt},
	every axis plot/.append style={line width=.9pt},
    every axis y label/.append style={yshift=-3pt}
}

\begin{document}

\ninept

\title{Nonnegative DAG Learning via Concomitant Estimation}

\name{Madeline Navarro$^1$, Gonzalo Mateos$^2$, and Samuel Rey$^3$ 
\thanks{This work was partially supported by the NSF grant 2231036, the Spanish AEI Grants PID2022-136887NB-I00 and PID2023-149457OB-I00, and by the Community of Madrid (via grants CAM-URJC F1180 (CP2301), TEC-2024/COM-89, and Madrid ELLIS Unit).Claude AI was used to assist in writing and coding the simulations. The authors take full responsibility for the results of this paper.
        Emails:   
        \href{mailto:m.navarro@northeastern.edu}{m.navarro@northeastern.edu}, 
        \href{mailto:gmateosb@ece.rochester.edu}{gmateosb@ece.rochester.edu},
        \href{mailto:samuel.rey.escudero@urjc.es}{samuel.rey.escudero@urjc.es}.
}}
\address{$^1$Dept. of Electrical and Computer Engineering, Northeastern University, Boston, Massachusetts, USA \\
$^2$Dept. of Electrical and Computer Engineering, University of Rochester, Rochester, New York, USA \\
$^3$Dept. of Signal Theory and Communications, King Juan Carlos University, Madrid, Spain}

\maketitle

\begin{abstract}
We study the problem of learning directed acyclic graphs (DAGs) with nonnegative edge weights from observational data. We propose the \emph{Nonnegative and Concomitant (NoCo) DAG estimator}, which jointly recovers the weighted graph structure and the exogenous noise variances in the linear structural equation model for the observations.
Different from prior art, this noise-adaptive formulation blends a smoothed concomitant lasso criterion with a simpler log-determinant acyclicity constraint that exploits nonnegativity and yields a more benign optimization landscape. Specifically, nonnegative weights allow us to impose acyclicity directly on the adjacency matrix without elementwise squaring of its entries, thus avoiding the well-documented degeneracy of the Karush-Kuhn-Tucker conditions.
Computationally, we develop a method of multipliers' algorithm that accommodates both homoscedastic and heteroscedastic noise profiles. Within each iteration, we use block successive convex approximation to minimize the augmented Lagrangian, alternating between proximal gradient steps for the adjacency matrix and closed-form updates for the noise scales. Simulated experiments demonstrate NoCo's improved structural and edge-weight recovery relative to competing methods in a variety of settings, highlighting the benefits of exploiting nonnegativity along with noise adaptivity.
\end{abstract}

\begin{keywords}
DAG learning, concomitant scale estimation, topology inference, causal discovery, constrained optimization.
\end{keywords}

\section{Introduction}
\label{S:intro}

Fields such as machine learning, statistics, and signal processing have long benefited from directed acyclic graphs (DAGs) as representations of various fundamental real-world systems~\cite{sachs2005causal,lucas2004bayesian}.
Notably, DAGs are widely employed to encode statistical dependencies in Bayesian networks for causal inference~\cite{peters2017elements,squires2023causal}.
Beyond probabilistic graphical models, hierarchical relationships are often represented as DAGs describing, e.g., feedforward neural network architectures, sequential tasks, ranked decision making processes~\cite{anselmi2026timeconstrained}, or even inductive biases in machine learning models~\cite{zhang2019dvae}.
The \emph{causal discovery} task of recovering DAGs from (passive) observational data is therefore critical for their adoption in practical settings.

Indeed, DAG structure learning continues to be actively researched as it is fraught with challenges~\cite{vowels2022d}.
Even with sufficient amounts of data, the task can be inherently ill-posed. This identifiability predicament arises when multiple candidate DAGs belong to the same Markov equivalence class (i.e., they induce the same set of conditional independencies as those in the data distribution)~\cite{loh2014highdimensional}.
To characterize performance or guarantee uniqueness, some methods require additional information, such as extra distributional assumptions or access to interventional manipulations~\cite{peters2014identifiability,xue2025dotears}. However, these may be unrealistic, resource-heavy, or unethical to implement.

Since DAG structure learning from observational data is in general an NP-hard problem~\cite{chickering2004large}, the associated computational challenges are well documented~\cite[Ch. 7]{peters2017elements}. Early score-based approaches required formidable discrete optimization due to the combinatorial acyclicity constraint; see e.g.,~\cite{chickering96np}.
Recent efforts advocate relaxing said discrete restriction using nonconvex, smooth functions of the DAG's adjacency matrix, whose zero level set coincides with the space of DAGs~\cite{zheng2018dags}.
This way, the search can be conducted using gradient-based \emph{continuous} optimization algorithms; a paradigm shift that led to significant innovations relative to the pioneering NOTEARS formulation~\cite{wei2020dags,bello2022dagma,ng2020role,saboksayr2024colide,rey2025nonnegative}.
Despite substantial progress, statistical and optimization challenges remain~\cite{reisach2021beware,ng2024structure}, and these nonconvex methods lack global optimality guarantees.

In addition to smooth acyclicity constraints, the choice of an appropriate score function (typically composed of data fitting and sparsity-promoting regularization terms) is key to guide first-order algorithms towards good-quality approximate solutions. A recent yet promising direction is to employ \emph{concomitant} estimators of scale parameters~\cite{saboksayr2024colide,mateos2026colide} inspired by the smooth concomitant lasso~\cite{smooth-con-lasso}, where the DAG adjacency matrix is estimated jointly with the exogenous noise variance in postulated linear structural equation models (SEMs).
Such \emph{noise-adaptivity} feature yields multiple benefits, such as mitigating the laborious hyperparameter tuning that is typically required because exogenous noise variances are unknown.
This also enables structure learning in heteroscedastic SEMs, where noise levels need not be identical across nodes; an implicit assumption of most existing methods~\cite{zheng2018dags,wei2020dags,bello2022dagma}. Like GOLEM~\cite{ng2020role}, concomitant DAG learning enjoys asymptotic recovery guarantees to a DAG quasi-equivalent to the ground-truth for general (i.e., heteroscedastic and hence nonidentifiable) linear Gaussian SEMs~\cite{mateos2026colide}.

To supplement these methodological advances, it is prudent to exploit favorable structure such as sparsity (of edges or root causes)~\cite{ng2020role,misiakos2023learning,mihal2026icassp}, or else knowledge about the topological ordering, a subset of edges, or the underlying distribution~\cite{ng2024structure,ajourlu2026build}.
In this work, we assume that the DAG of interest has \emph{nonnegative edge weights}. This covers many realistic scenarios, such as edges representing excitatory causal effects, binary indicators of pairwise influence, or stage transitions in task scheduling problems~\cite{peters2017elements,anselmi2026timeconstrained}.
Imposing nonnegativity could also be a convenient design choice for subsequent construction of meaningful orthogonal Fourier bases~\cite{mihal2026icassp}.
Our main contribution is a novel Nonnegative and Concomitant (NoCo) DAG estimator blending the merits of concomitant learning and nonnegativity~\cite{mateos2026colide,rey2026exploiting} to unlock friendlier optimization, recovery guarantees, and effective as well as robust performance that we demonstrate via numerical experiments.

\section{Fundamentals of DAG learning}
\label{S:bg}

Here we review the necessary background on DAGs, linear SEMs, and structure identification methods from nodal observations. These fundamentals allow us to formally state our DAG learning problem.

\medskip\noindent{\bf Directed acyclic graphs.}
We let $\ccalG = (\ccalV,\ccalE)$ denote a graph of $N$ nodes $\ccalV$ with \emph{directed} edges $\ccalE \subseteq \ccalV \times \ccalV$, where $(i,j) \in \ccalE$ represents the link from node $i$ to node  $j$.
The structure of $\ccalG$ can be conveniently represented by the adjacency matrix $\bbA \in \reals^{N \times N}$, where $A_{ij}$ denotes the weight of the edge $(i,j)$, with $A_{ij}\neq 0$ if and only if $(i,j) \in \ccalE$.
For such a directed graph $\ccalG$ to be a DAG, there must be no cycles or self-loops in $\ccalG$.
In such a case, we say that $\bbA \in \mbD$, where $\mbD$ is the set of valid adjacency matrices corresponding to DAGs.
In this work, we assume $\ccalG$ is not only a DAG but also has \emph{nonnegative} edge weights, so $\bbA \in \reals_+^{N\times N}$.
We will show how nonnegativity leads us to a friendlier optimization environment.

\medskip\noindent{\bf Linear SEMs.}
DAGs are often integral parts of probabilistic graphical models, where our set of nodes $\ccalV$ comprise the random vector $\bbx = [x_1,\dots,x_N]^\top$, and the edges $\ccalE$ encode conditional independencies between node pairs.
In Markovian models with respect to a given DAG, each random variable $x_i$ is dependent only on its parents ${\rm PA}_i = \{ j \in \ccalV : A_{ji} \neq 0 \}$ and is conditionally independent of all other nodes~\cite{mateos2026colide}.
We consider one of the most common dependencies between parent and child nodes, linear SEMs, where $x_i = \sum_{j \in {\rm PA}_i} A_{ji} x_j + z_i$ for every $i \in \{1,\dots,N\}$, with $\bbz = [z_1,\dots,z_N]^\top$ representing mutually independent, exogeneous noise~\cite{peters2017elements}.
When given a collection of $M$ i.i.d. samples $\bbX = [\bbx_1,\dots,\bbx_M]\in\reals^{N\times M}$ with corresponding noise $\bbZ = [\bbz_1,\dots, \bbz_M]\in\reals^{N\times M}$, we may write the linear SEM as 
\begin{equation}\label{eq:sem_signals}
    \bbX = \bbA^\top\bbX + \bbZ.
\end{equation}
We let $\bbsigma = [\sigma_1,\dots,\sigma_N]^\top$ collect the standard deviations of the entries in $\bbz$. When $\sigma_1=\dots=\sigma_N$, the noise is homoscedastic. In contrast, when the variances differ across nodes, the noise is heteroscedastic.
In either case, the standard deviation matrix of $\bbz$ is $\bbSigma := \diag(\bbsigma)$, with the entries of $\bbsigma$ populating the diagonal.

\medskip\noindent{\bf DAG structure learning.}
Given a set of observations $\bbX$, the goal of DAG structure learning is to recover the underlying DAG encoded by $\bbA$.
\emph{Score-based} approaches seek an estimate $\hbA$ by minimizing a loss function $F(\bbA) = F(\bbA;\bbX)$ that measures how well a candidate graph fits the observations.
However, the estimated matrix must also satisfy $\hbA \in \mbD$ to ensure that it represents a valid DAG.
Since this constraint is notoriously difficult to enforce, a popular approach replaces it with a smooth acyclicity constraint, yielding the optimization problem
\alna{
    \hbA \in
    \argmin_{\bbA}
    ~
    F(\bbA)
    \quad{\rm s.t.}\quad
    h(\bbA) = 0.
\label{eq:vanilla_dag_learning}}
The \emph{acyclicity condition} $h(\bbA) = 0$ is equivalent to the explicit constraint $\bbA \in \mbD$, with $h:\reals^{N\times N} \rightarrow \reals$ as a nonconvex, smooth function whose zero level set is $\mbD$~\cite{zheng2018dags}.
The seminal work that introduced this concept proposed the NOTEARS mapping $h_{\rm NT} = \tr(e^{\bbA \circ \bbA}) - N$, where $\circ$ denotes the Hadamard product~\cite{zheng2018dags}.
NOTEARS advocates a very natural function for~\eqref{eq:vanilla_dag_learning}, as it sums the diagonal entries of powers of $\bbA\circ\bbA$, with the magnitudes of $\diag(\bbA\circ\bbA)^k$ proportional to the number of $k$-cycles per node.
Adaptations of $h_{\rm NT}$ follow its example of ensuring only DAGs in the zero level set with modifications to improve the optimization landscape.
Most germane to our work is the log-determinant variant in DAGMA~\cite{bello2022dagma}, i.e., $h_{\rm DM}(\bbA;u) = N \log u - \log\det(u \bbI - \bbA \circ \bbA)$, with $u \geq 0$ ensuring a well-defined mapping. Without loss of generality, we let $u = 1$ for $h_{\rm DM}(\bbA) = h_{\rm DM}(\bbA;1)$, which is empirically observed to perform well when restricted to adjacency matrices with spectral radius $\rho(\bbA) < 1$ (this contains $\mbD$, since DAGs have $\rho(\bbA)=0$).

While NOTEARS and its offshoots constitute a major development in DAG learning, the problem~\eqref{eq:vanilla_dag_learning} still suffers obstacles.
First, these methods score candidate DAGs via an ordinary least-squares criterion, which implicitly assumes identical exogenous noise variances across nodes.
Departures from this assumption bias structure recovery~\cite{loh2014highdimensional}, rendering such methods ill-suited to heteroscedastic settings.
Second, the composition of non-convex functions with the Hadamard product leads to an adverse optimization landscape with limited guarantees~\cite{reisach2021beware,ng2022convergence,ng2024structure}.
To mitigate these challenges, our method leverages a noise-adaptive score function while exploiting nonnegative edge weights for a more benign optimization landscape, as detailed in the next section.

\section{Nonnegative and concomitant\\ DAG learning}
\label{S:method}

Our goal is to jointly estimate the target DAG $\bbA^*$ and the noise levels $\bbsigma^*$, assuming that $\bbA^*$ has nonnegative edge weights.
To this end, we propose the Nonnegative and Concomitant (NoCo) DAG estimator, which builds on concomitant estimation~\cite{saboksayr2024colide,mateos2026colide} and exploits nonnegativity to simplify the acyclicity function~\cite{rey2025nonnegative,rey2026exploiting}.
This approach involves solving the following optimization problem
\alna{
    &\min_{\bbA \geq \bbzero, \bbsigma \geq \bbsigma_0}&
    ~
    F(\bbA, \bbsigma )
    ~\,
    {\rm s.t.}
    ~\,
    h(\bbA) = 0,
    ~
    \rho(\bbA) < 1.
\label{eq:noco_dag_learning}}
Although the formulation can accommodate any acyclicity function $h$, we focus on log-determinant-based functions similar to $h_{\rm DM}$.
For these functions, the spectral radius constraint ensures that the log-determinant is well defined.
The inequalities are applied elementwise, with $\bbA \geq \bbzero$ ensuring nonnegative edge weights and $\bbsigma \geq \bbsigma_0$ precluding ill-posed scenarios with zero-valued variances.
Following~\cite{saboksayr2024colide,mateos2026colide}, we set $[\bbsigma_0]_i := 10^{-2}\sqrt{\hat{C}_{ii}}$ for each $i \in \{1,\dots,N\}$, where $\hbC := \frac{1}{M}\bbX\bbX^\top$ is the sample covariance matrix.
Notably, our noise-adaptive fidelity loss $F(\bbA,\bbsigma)$ incorporates the exogenous noise levels $\bbsigma$. For the general heteroscedastic setting, this loss is given by
\alna{
    F(\bbA,\bbsigma)
    &~:=~&
    {\textstyle\frac{1}{2M}}
    \tr
    \big(
        \bbX^\top
        (\bbI - \bbA)^{\top}
        \diag(\bbsigma)^{-1}
        (\bbI - \bbA)
        \bbX
    \big)
&\nonumber\\&
    && \qquad
    +
    {\textstyle\frac{1}{2}}
    \bbone^\top \bbsigma
    +
    \alpha
    \bbone^\top \bbA \bbone,
\label{eq:fidelity}}
where the third term is equivalent to the popular $\ell_1$-norm sparsity penalty $\|\bbA\|_1$ under our assumption of nonnegativity.
A key advantage of this score function is that it is jointly convex in $\bbA$ and $\bbsigma$, so the nonconvexity of~\eqref{eq:noco_dag_learning} stems entirely from the acyclicity and spectral radius constraints.
Although the formulation is presented for heteroscedastic noise, it readily accommodates the homoscedastic case by imposing $\bbsigma = \sigma\bbone$, where $\sigma$ is a shared scalar noise parameter.

Beyond converting the sparse regularizer into a linear penalty, specializing the problem to nonnegative edges offers particular advantages for DAG structure learning~\cite{rey2025nonnegative,rey2026exploiting}.
Since we need not account for cycles in $\bbA$ whose weights sum to 0, we can apply popular acyclicity functions without the Hadamard product $\bbA \circ \bbA$.
This simplification yields a more favorable optimization landscape and allows for local recovery guarantees for $\bbA^*$ in the population regime~\cite{bello2022dagma,rey2026exploiting}.
Due to its theoretical and practical advantages, we consider the NOMAD modification of the acyclicity function $h_{\rm DM}$ over the domain $\mbA_+ := \big\{ \bbA \in \reals_+^{N \times N} : \rho(\bbA) < 1 \big\}$~\cite{rey2025nonnegative,rey2026exploiting},
\alna{
    h_{\rm NM}(\bbA)
    &~:=~&
    -\log \det(\bbI - \bbA),
\label{eq:acyclicity_nomad}}
hence the need for the spectral radius constraint in~\eqref{eq:noco_dag_learning}.

\subsection{Algorithmic implementation}
Absent the Hadamard product, the function in~\eqref{eq:acyclicity_nomad} avoids the degeneracy of the Karush--Kuhn--Tucker (KKT) conditions associated with standard acyclicity functions such as $h_{\rm NT}$ and $h_{\rm DM}$~\cite{wei2020dags,rey2026exploiting}.
Solving~\eqref{eq:noco_dag_learning} is therefore amenable to the method of multipliers, an iterative algorithm based on minimizing the augmented Lagrangian that provides a principled way to solve equality-constrained optimization problems~\cite[Ch. 4.2]{bertsekas2016nonlinear}.
Here, the augmented Lagrangian is given by
\alna{
    L_t(\bbA,\bbsigma)
    :=
    F\big(\bbA,\bbsigma\big)
    +
    \lambda_t
    h(\bbA)
    +
    \frac{c_t}{2}
    h^2(\bbA),
\label{eq:aug_lag}}
where $\lambda_t,c_t > 0$ respectively denote the Lagrange multiplier and the penalty parameter, and $t\in\naturals$ denotes the iteration index.

NoCo is tabulated under Algorithm~\ref{alg:noco}. Following the method of multipliers recipe, each outer iteration comprises three steps: a primal update, a dual update, and a penalty parameter update.

\noindent\textbf{Step 1 - Primal update.}
For fixed $\lambda_t$ and $c_t$, the primal update is
\begin{equation}
    \big(\bbA^{(t+1)},\bbsigma^{(t+1)}\big)
    \in \arg\min_{\substack{\bbA \geq \bbzero,\; \bbsigma \geq \bbsigma_0\\ \rho(\bbA)<1}}
    L_t(\bbA,\bbsigma).
    \label{eq:primal_subproblem}
\end{equation}
This optimization problem can be tackled in several ways. 
Here, we benefit from the benign structure of the score function and approximately solve it using block successive convex approximation (SCA)~\cite{yang2020inexact, saboksayr2024block}.
SCA is an iterative method that updates each block of variables by minimizing a convex surrogate of the objective while holding the other blocks fixed. 
Starting from $\bbB^{(0)}=\bbA^{(t)}$ and $\bbs^{(0)}=\bbsigma^{(t)}$, we alternate DAG and noise scale updates.
Specifically, at iteration $j$, we fix $\bbs^{(j)}$ and update the DAG minimizing the quadratic approximation of $L_t$ given by
\alna{
    &~\min_{\bbA}~&
    \Big\langle
        \nabla_{\bbA}
        L_t\big(
            \bbB^{(j)}, \bbs^{(j)}
        \big),
        \bbA - \bbB^{(j)}
    \Big\rangle
    +
    \frac{1}{2\eta_{t,j}}
    \big\|
        \bbA - \bbB^{(j)}
    \big\|_F^2
&\nonumber\\&
    &~{\rm s.t.}~&
    ~~
    \bbA \geq \bbzero, ~~
    \rho(\bbA) < 1,
\label{eq:A_update}}
Choosing a stepsize $\eta_{t,j}>0$ to keep the update within the feasible set detrmined by the constraint $\rho(\bbA)<1$, the solution of~\eqref{eq:A_update} is obtained via the projected gradient step
\begin{equation}
    \bbB^{(j+1)} = \Big[
        \bbB^{(j)} - \eta_{t,j}
        \nabla_{\bbA}L_t\big(\bbB^{(j)},\bbs^{(j)}\big)
    \Big]_+,
    \label{eq:A_prox_update}
\end{equation}
where $[\cdot]_+$ denotes the elementwise projection onto the nonnegative orthant.

Regarding the noise parameter update, since the acyclicity terms do not depend on $\bbsigma$, minimizing~\eqref{eq:primal_subproblem} with respect to $\bbsigma$ for fixed $\bbB^{(j+1)}$ reduces to a convex problem.
Furthermore, its exact solution is given in closed form by
\alna{
    s_i^{(j+1)}
    &\,=\,&
    \sqrt{\max\!
    \Big( 
        \big[
            (\bbI - \bbB^{(j+1)})
            \hbC
            (\bbI - \bbB^{(j+1)})^{\top}
        \big]_{ii}
        , \,
        [\bbsigma_0]_i
    \Big)
    }
&\nonumber\\&
    &&
    \qquad\qquad\qquad\qquad\qquad
    \forall~i \in \{1,\dots,N\}.
\label{eq:sigma_update}}
These updates are repeated until the inner stopping criterion is met, after which the final inner iterates define $\bbA^{(t+1)}$ and $\bbsigma^{(t+1)}$.

\noindent\textbf{Step 2 - Dual update.}
Let $h_t:=h_{\rm NM}(\bbA^{(t)})$ denote the acyclicity residual.
The Lagrange multiplier is updated according to
\begin{equation}
    \lambda_{t+1}=\lambda_t+c_{t}h_{t+1}.
    \label{eq:lambda_update}
\end{equation}
This update can also be interpreted as a dual gradient ascent step, since the constraint violation $h(\bbA^{(t+1)})$ corresponds to the gradient of the augmented Lagrangian with respect to $\lambda$.

\noindent\textbf{Step 3 - Penalty parameter update.}
The penalty parameter is increased when the acyclicity residual has not decreased sufficiently, following
The penalty parameter increases when the acyclicity residual has not decreased sufficiently, following
\begin{equation}
    c_{t+1}=\begin{cases}
        \beta c_t, & h_{t+1}>\gamma h_t,\\
        c_t, & \text{otherwise},
    \end{cases}
    \qquad \beta>1,\quad \gamma\in(0,1).
    \label{eq:c_update}
\end{equation}
The parameter $\gamma$ specifies the required reduction in the residual, while $\beta$ controls the increase in the penalty.
Observe that as the range of $h_{\rm NM}$ is nonnegative over $\mbA_+$, every $h_t$ is nonnegative, so $c_{t+1}$ is monotonically nondecreasing and $\lambda_{t+1}$ increasing in $t \in \naturals$.

\renewcommand{\algorithmicrequire}{\textbf{Input:}}
\renewcommand{\algorithmicensure}{\textbf{Output:}}

\algsetup{indent=2em, linenosize=\small}
\begin{figure}[t]
    \centering
    \scalebox{.83}{
    \begin{minipage}{.56\textwidth}
        \vspace{-.2cm}
    \begin{algorithm}[H]
        \caption{Nonnegative and Concomitant (NoCo) DAG Estimation}
        \label{alg:noco}
    \begin{algorithmic}[1]
        \REQUIRE 
            Data 
            $\bbX$, 
            hyperparameters 
            $\bbsigma_0$, 
            $\alpha \geq 0$, 
            $\beta > 1$, 
            $\gamma \in (0,1)$
        \STATE
        Initialize 
        $\bbA^{(0)} = \bbzero$, ~
        $\bbsigma^{(0)} = \bbsigma_0$, ~
        $\lambda_0$, ~
        $c_0$, ~
        $h_0 = h_{\rm NM}(\bbA^{(0)})$, ~
        $t = 0$.
        \WHILE{Outer stopping criterion not met}
            \STATE 
            {\bf Primal update:} 
                Set $\bbB^{(0)} = \bbA^{(t)}$,
                $\bbs^{(0)} = \bbsigma^{(t)}$,
                and $j = 0$.
            \WHILE{Inner stopping criterion not met}
                \STATE Choose $\eta_{t,j}>0$ so that the graph update satisfies $\rho(\bbB^{(j+1)})<1$.
                \STATE
                Obtain 
                    $\bbB^{(j+1)}$ via~\eqref{eq:A_prox_update} and 
                    $\bbs^{(j+1)}$ via~\eqref{eq:sigma_update}.
                \STATE
                Update $j \leftarrow j + 1$.
            \ENDWHILE
            \STATE
                Set $\bbA^{(t+1)} = \bbB^{(j)}$ and 
                $\bbsigma^{(t+1)} = \bbs^{(j)}$.
            \STATE 
            {\bf Dual update:} Compute $h_{t+1}=h_{\rm NM}(\bbA^{(t+1)})$ and update $\lambda_{t+1}$ via~\eqref{eq:lambda_update}.
            \STATE
            {\bf Penalty parameter update:} Update $c_{t+1}$ via~\eqref{eq:c_update}.
            \STATE 
            Update $t \leftarrow t + 1$.
        \ENDWHILE
        \ENSURE DAG $\hbA = \bbA^{(t)}$ and noise variances $\hat{\bbsigma} = \bbsigma^{(t)}$
    \end{algorithmic}
    \end{algorithm}
    \end{minipage}
    }
\vspace{-.3cm}
\end{figure}

\begin{figure*}[t!]
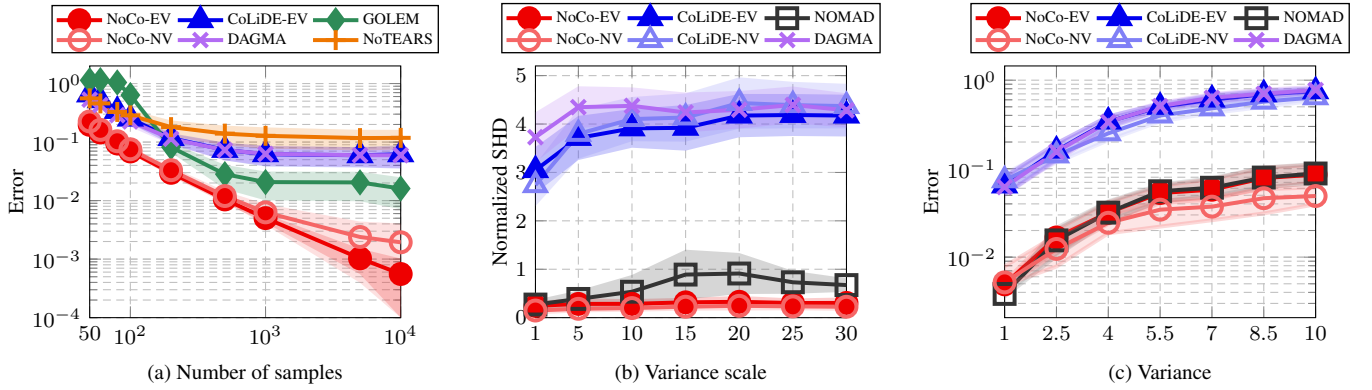

    \centering
    \tikzset{every mark/.append style={scale=1}}
    \pgfplotsset{
        legend style={font=\fontsize{6}{7}\selectfont},
        xlabel style={at={(axis description cs:0.5,-0.15)},anchor=north},
    }
    \begin{minipage}[t]{.32\textwidth}
        \begingroup
\input{figures/experiment_plot_styles.tex}
\begin{tikzpicture}[baseline,scale=1]
\pgfplotstableread[col sep=comma]{data/samples_hom_error.csv}\sampletable
\begin{loglogaxis}[
    experiment axis,
    xlabel={(a) Number of samples},ylabel={Error},
    xmin=50,xmax=10000, ymin=0.0001,ymax=2,
    xtick={50,100,1000,10000},
    xticklabels={$50$,$10^2$,$10^3$,$10^4$},
    ytick={0.0001,0.001,0.01,0.1,1},
]
\experimentcurve{ncev}{NoCo-EV}{0.0001}{\sampletable}
\experimentcurve{coev}{CoLiDE-EV}{0.0001}{\sampletable}
\experimentcurve{golem}{GOLEM}{0.0001}{\sampletable}
\experimentcurve{ncnv}{NoCo-NV}{0.0001}{\sampletable}
\experimentcurve{dagma}{DAGMA}{0.0001}{\sampletable}
\experimentcurve{notears}{NoTEARS}{0.0001}{\sampletable}
\end{loglogaxis}
\end{tikzpicture}
\endgroup
    \end{minipage}\hfill
    \hspace{.6cm}
    \begin{minipage}[t]{.32\textwidth}
        \begingroup
\input{figures/experiment_plot_styles.tex}
\begin{tikzpicture}[baseline,scale=1]
\pgfplotstableread[col sep=comma]{data/variance_hetero_shd.csv}\variancetable
\begin{axis}[
    experiment axis,
    xlabel={(b) Variance scale},ylabel={Normalized SHD},
    xmin=1,xmax=30,ymin=0,ymax=5.2,
    xtick={1,5,10,15,20,25,30},
    ytick={0,1,2,3,4,5},
]
\experimentcurve{ncev}{NoCo-EV}{0}{\variancetable}
\experimentcurve{coev}{CoLiDE-EV}{0}{\variancetable}
\experimentcurve{nomad}{NOMAD}{0}{\variancetable}
\experimentcurve{ncnv}{NoCo-NV}{0}{\variancetable}
\experimentcurve{conv}{CoLiDE-NV}{0}{\variancetable}
\experimentcurve{dagma}{DAGMA}{0}{\variancetable}
\end{axis}
\end{tikzpicture}
\endgroup
    \end{minipage}\hfill
    \begin{minipage}[t]{.32\textwidth}
        \begingroup
\begin{tikzpicture}[baseline,scale=1]

\pgfplotstableread{data/shd_data.csv}\shdtable
\pgfplotstableread{data/err_data.csv}\errtable

\pgfmathsetmacro{\opacity}{0.2}
\pgfmathsetmacro{\contourop}{0.25}


\begin{semilogyaxis}[
        xlabel={(c) Variance},
        xmin=1,
        xmax=10,
        xtick={1,2.5,4,5.5,7,8.5,10},
        ylabel={Error},
        every axis plot/.append style={line width=1.5pt},
        grid style=densely dashed,
        grid=both,
        legend style={
            at={(0.5, 1.05)},
            anchor=south},
        legend columns=3,
        width=\linewidth,
    ]

    \addplot[
        white!40!Red2, name path = ncev-lo, opacity=\contourop, forget plot
    ] table [
        x=x, y=ncev_lo
    ] \errtable;
    \addplot[
        white!40!Red2, name path = ncev-hi, opacity=\contourop, forget plot
    ] table [
        x=x, y=ncev_hi
    ] \errtable;
    \addplot[
        white!30!Red2, fill opacity=\opacity, forget plot
    ] fill between[of=ncev-lo and ncev-hi];
    \addplot[
        Red2, solid, mark=*, mark size=2.5pt
    ] table [
        x=x, y=ncev_agg
    ]{\errtable};

    \addplot[
        white!30!Blue2, name path = coev-lo, opacity=\contourop, forget plot
    ] table [
        x=x, y=coev_lo
    ] \errtable;
    \addplot[
        white!30!Blue2, name path = coev-hi, opacity=\contourop, forget plot
    ] table [
        x=x, y=coev_hi
    ] \errtable;
    \addplot[
        white!30!Blue2, fill opacity=\opacity, forget plot
    ] fill between[of=coev-lo and coev-hi];
    \addplot[
        Blue2, solid, mark=triangle*, mark size=3pt
    ] table [
        x=x, y=coev_agg
    ]{\errtable};

    \addplot[
        black!50!white, name path = nomad-lo, opacity=\contourop, forget plot
    ] table [
        x=x, y=nomad_lo
    ] \errtable;
    \addplot[
        black!50!white, name path = nomad-hi, opacity=\contourop, forget plot
    ] table [
        x=x, y=nomad_hi
    ] \errtable;
    \addplot[
        black!50!white, fill opacity=\opacity, forget plot
    ] fill between[of=nomad-lo and nomad-hi];
    \addplot[
        black!80!white, solid, mark=square, mark size=2.5pt
    ] table [
        x=x, y=nomad_agg
    ]{\errtable};

    \addplot[
        white!40!Red2, name path = ncnv-lo, opacity=\contourop, forget plot
    ] table [
        x=x, y=ncnv_lo
    ] \errtable;
    \addplot[
        white!40!Red2, name path = ncnv-hi, opacity=\contourop, forget plot
    ] table [
        x=x, y=ncnv_hi
    ] \errtable;
    \addplot[
        white!50!Red2, fill opacity=\opacity, forget plot
    ] fill between[of=ncnv-lo and ncnv-hi];
    \addplot[
        white!40!Red2, solid, mark=o, mark size=2.5pt
    ] table [
        x=x, y=ncnv_agg
    ]{\errtable};

    \addplot[
        white!70!Blue2, name path = conv-lo, opacity=\contourop, forget plot
    ] table [
        x=x, y=conv_lo
    ] \errtable;
    \addplot[
        white!70!Blue2, name path = conv-hi, opacity=\contourop, forget plot
    ] table [
        x=x, y=conv_hi
    ] \errtable;
    \addplot[
        white!70!Blue2, fill opacity=\opacity, forget plot
    ] fill between[of=conv-lo and conv-hi];
    \addplot[
        white!50!Blue2, solid, mark=triangle, mark size=3pt
    ] table [
        x=x, y=conv_agg
    ]{\errtable};

    \addplot[
        white!30!Purple2, name path = dagma-lo, opacity=\contourop, forget plot
    ] table [
        x=x, y=dagma_lo
    ] \errtable;
    \addplot[
        white!30!Purple2, name path = dagma-hi, opacity=\contourop, forget plot
    ] table [
        x=x, y=dagma_hi
    ] \errtable;
    \addplot[
        white!30!Purple2, fill opacity=\opacity, forget plot
    ] fill between[of=dagma-lo and dagma-hi];
    \addplot[
        white!30!Purple2, solid, mark=x, mark size=2.5pt
    ] table [
        x=x, y=dagma_agg
    ]{\errtable};

    \legend{
        NoCo-EV,
        CoLiDE-EV,
        NOMAD,
        NoCo-NV,
        CoLiDE-NV,
        DAGMA
    }

\end{semilogyaxis}
\end{tikzpicture}
\endgroup
    \end{minipage}
    \vspace{-.3cm}
    \caption{Comparison of DAG-learning methods: (a) estimation error as sample size increases under homoscedastic noise; (b) normalized SHD versus noise variance under heteroscedastic noise across all nodes; (c) estimation error versus noise variance on a subset of nodes, with fixed variance in the remaining ones. Results show the mean and standard deviation over 100 realizations.}
    \label{fig:experiments}
\end{figure*}

The proposed NoCo algorithm has a computational complexity of $\ccalO(N^3)$ and it enjoys particular advantages due to the combination of concomitant learning and nonnegative edge weights.
First, our approach builds on concomitant DAG estimation, for which asymptotic quasi-equivalence guarantees have been established under suitable assumptions~\cite{saboksayr2024colide,mateos2026colide}.
Separately, under mild step-size conditions, the block SCA iterations in Step~1 converge to a stationary point of the primal subproblem~\cite{yang2019inexact}.
Second, $h_{\rm NM}$ avoids the KKT degeneracy of Hadamard-based functions such as $h_{\rm DM}$~\cite{rey2026exploiting}, which is more impactful for concomitant learning since DAG stationarity $\bbB^{(j+1)} = \bbB^{(j)}$ implies $\bbs^{(j+1)} = \bbs^{(j)}$.
More specifically, $\nabla_{\bbA}h_{\rm DM}(\bbA) = \bbzero$ for any $\bbA \in \mbD$, but we may have $\nabla_{\bbA}h_{\rm NM}(\bbA) \neq \bbzero$ even if $\bbA \in \mbD$. 
Furthermore, if $\bbB^{(j)} \in \mbD$ , then the update in~\eqref{eq:A_update} will exploit both the fidelity loss $F$ and the acyclicity function $h_{\rm NM}$ for obtaining the udpated $\bbB^{(j+1)}$, whereas if we let $h = h_{\rm DM}$ in~\eqref{eq:noco_dag_learning}, then~\eqref{eq:A_update} would only account for $F$, even though we cannot guarantee that perturbing $\bbB^{(j)}$ in the direction of $\nabla_{\bbA}F(\bbB^{(j)},\bbs^{(j)})$ will still be a DAG.
Third, if we initialize small values of $\lambda_0$ and $c_0$, the monotonicity of $\lambda_t$ and $c_t$ with respect to $t \in \naturals$ implements a type of automatic scheduling.
In initial iterations, Algorithm~\ref{alg:noco} emphasizes data fidelity and noise level estimation.
Then, as the value of $F(\bbA^{(t)}, \bbsigma^{(t)})$ reduces and $\lambda_t, c_t$ increase, the updates then promote DAG validity and discourage cycles in $\bbA^{(t)}$.
As a final remark on practicality, note that~\eqref{eq:A_update} is amenable to acceleration, which we implement experimentally in the next section via Adam updates~\cite{kingma2015adam}.
Finally, since the score function in~\eqref{eq:fidelity} rescales the residuals, the sparsity parameter $\alpha$ is decoupled from the noise scale, with minimax-optimal scaling $\alpha \asymp \sqrt{\log N/M}$~\cite{mateos2026colide}.

\section{Experiments}
\label{S:experiments}

We evaluate NoCo in comparison with existing DAG learning methods across various scenarios to demonstrate its advantages.
We aim to recover a target DAG $\bbA^*$ via an estimated $\hbA$, which we assess via the structural Hamming distance (SHD) and normalized Frobenius error $\| \hbA - \bbA^* \|_F^2 / \| \bbA^* \|_F^2$.
We compare our method, denoted ``{\bf NoCo}'', to relevant existing works: ``{\bf NoTEARS}''~\cite{zheng2018dags}, ``{\bf DAGMA}''~\cite{bello2022dagma}, ``{\bf GOLEM}''~\cite{ng2020role}, ``{\bf NOMAD}''~\cite{rey2026exploiting}, and two versions of CoLiDE~\cite{saboksayr2024colide}: ``{\bf CoLiDE-EV}'', which assumes homoscedasticity $\bbsigma^* = \sigma^*\bbone$, and ``{\bf CoLiDE-NV}'', which permits heteroscedasticity.
Unless stated otherwise, we generate Erd\H{o}s-R\'{e}nyi (ER) DAGs of $N = 100$ nodes with an average node degree of $4$ and $M = 10^3$ samples following the linear SEM introduced in \eqref{eq:sem_signals} with standard Gaussian exogeneous noise\footnote{\url{https://github.com/reysam93/nonneg_colide}}.\vspace{2pt}

\noindent \textbf{Test case 1 - Sample size.}
We first examine how estimation performance varies with sample size in the identifiable homoscedastic setting, varying $M$ from $50$ to $10^4$.
Fig.~\ref{fig:experiments}(a) reports the estimation error, where we observe that both NoCo variants consistently outperform all other baselines.
More interestingly, although all algorithms benefit from additional samples, the improvements diminish for those methods agnostic to the nonnegativity of the target DAG.
This contrasts with the behaviour of NoCo, whose errors continue to decrease as more samples become available, highlighting the benefits of employing a simpler acyclicity function that harnesses the nonnegativity of the DAG.
Furthermore, as expected, NoCo-EV outperforms NoCo-NV, in agreement with its equal-variance assumption matching the noise model used in this experiment.\vspace{2pt}

\noindent \textbf{Test case 2 - Heteroscedastic noise.}
Next, we study the impact of increasing the noise variance in the heteroscedastic regime.
In this experiment, we first draw the nodewise noise standard deviations uniformly from $[0.5,5]$, then normalize the corresponding variances to match the desired scale.
Fig.~\ref{fig:experiments}(b) shows the mean SHD normalized by the number of nodes as the scale of the variance increases.
The results illustrate how NoCo and NOMAD, which exploit the nonnegativity of the DAG, substantially outperform the alternatives that ignore this structure.
In particular, NoCo-NV achieves the lowest SHD throughout the tested range, demonstrating accurate graph recovery despite being in a challenging setting.
Additional results show that its scale-invariant error in estimating the noise standard deviation profile remains below $0.03$.
These observations suggest that jointly estimating the graph and the relative noise levels encoded in $\bbSigma$ may help mitigate the challenges posed by heteroscedastic noise.
This promising behavior points to an interesting research direction for future work.\vspace{2pt}

\noindent \textbf{Test case 3 - Heteroscedastic noise on a subset of nodes.}
Finally, we assess robustness to increased noise levels localized to a subset of nodes.
To this end, we select $30\%$ of the nodes uniformly at random and increase their noise variance from $1$ to $10$, while keeping it fixed at $1$ on all remaining nodes.
Fig.~\ref{fig:experiments}(c) reports the estimation error as a function of the noise variance on this subset.
Consistent with the previous experiments, NoCo and NOMAD outperform the methods that do not exploit nonnegativity.
When all noise variances equal 1, we recover the homoscedastic setting for which NOMAD is tailored, and this baseline achieves the lowest error.
However, as the variance on the selected nodes increases, NoCo-NV attains the best performance in every tested heteroscedastic setting, reducing the mean estimation error by $44.4\%$ relative to NOMAD when this variance reaches 10.


\section{Conclusion}
\label{S:conclusion}

In this work, we presented NoCo, a novel approach to learning DAGs with nonnegative edge weights while jointly estimating exogenous noise variances.
Our framework marries the methodological advantages of concomitant learning and the optimization benefits arising from having nonnegative DAGs.
In particular, the robustness and recovery guarantees of estimating noise levels with the DAG of interest carry over to our approach.
Moreover, nonnegative edge weights permit simpler, more meaningful acyclicity functions.
This yields a friendlier optimization landscape, which is more impactful for concomitant learning since it affects algorithmic updates to both the DAG structure and noise variances.
Our experiments illustrate the benefits of this combination across different noise settings.
As nonnegativity introduces simple yet elegant structure, a natural and important future direction is investigating recovery and convergence guarantees for concomitant methods in this setting.

\section{Compliance with Ethical Standards}

The experiments in this paper are numerical simulation studies for which no ethical approval was required.


\bibliographystyle{IEEEbib}
\bibliography{citations}

\end{document}